\documentclass[sigconf,nonacm]{acmart}
\usepackage{multirow}

\AtBeginDocument{%
  \providecommand\BibTeX{{%
    \normalfont B\kern-0.5em{\scshape i\kern-0.25em b}\kern-0.8em\TeX}}}

\begin{document}

% \title[DeepImpact: Quantifying the Impact of Deep Learning for Computer Science Domain using Causal Inference]{DeepImpact: Quantifying the Impact of Deep Learning \\for Computer Science Domain using Causal Inference}
\title{A Causal Inference Approach for Quantifying Research Impact}


\author{Keiichi Ochiai}
\affiliation{%
  \institution{The University of Tokyo}
  \country{Japan}
}
\authornote{Concact author: ochiai@weblab.t.u-tokyo.ac.jp}

\author{Yutaka Matsuo}
\affiliation{%
  \institution{The University of Tokyo}
  \country{Japan}
}


\begin{abstract}
Deep learning has had a great impact on various fields of computer science by enabling data-driven representation learning in a decade.
Because science and technology policy decisions for a nation can be made on the impact of each technology,
quantifying research impact is an important task.
The number of citations and impact factor can be used to measure the impact for individual research.
What would have happened without the research, however, is fundamentally a counterfactual phenomenon.
Thus, we propose an approach based on causal inference to quantify the research impact of a specific technical topic.
We leverage difference-in-difference to quantify the research impact by applying to bibliometric data.
First, we identify papers of a specific technical topic using keywords or category tags from Microsoft Academic Graph, which is one of the largest academic publication dataset.
Next, we build a paper citation network between each technical field.
Then, we aggregate the cross-field citation count for each research field.
Finally, the impact of a specific technical topic for each research field is estimated by applying difference-in-difference.
Evaluation results show that deep learning significantly affects computer vision and natural language processing. Besides, deep learning significantly affects cross-field citation especially for speech recognition to computer vision and natural language processing to computer vision.
Moreover, our method revealed that the impact of deep learning was 3.1 times of the impact of interpretability for ML models.
\end{abstract}


\maketitle

\section{Introduction}
% 1.What is the problem?
Deep learning made a remarkable impact on many scientific disciplines especially in computer science domain such as image recognition and natural language processing ~\cite{lecun2015deep}.
Deep Learning has been used in various fields since the success of image recognition at the ImageNet Large Scale Visual Recognition Challenge (ILSVRC) in 2012~\cite{krizhevsky2012imagenet}. 
As a result, Yoshua Bengio, Geoffrey Hinton, and Yann LeCun received the 2018 ACM A.M. Turing Award for breakthroughs of deep neural networks that revolutionalized in Artificial Intelligence (AI)\footnote{https://awards.acm.org/about/2018-turing}.
Although deep learning made a qualitatively significant impact on AI,
a quantitative impact remains unveiled.
It is important to quantify the impact of research on a technical topic, not only deep learning, on the research field. 
Therefore, we aim to consider how to quantify the impact of research on a technical topic.

% 2. Why is it interesting and important?
Quantifying research impact is an important task for a national government and individual researchers.
From the perspective of a national government, in national science and technology policy, it is necessary to measure the impact of research because the national government plans its budget and personnel based on the prediction of technological development.
From the perspective of individual researchers, it is useful to understand the influence of past research when considering research themes.

% 3. Why is it hard? (E.g., why do naive approaches fail?)
A naive method to quantify the research impact is to measure the number of citations of individual research. 
When measuring the effects of a technical topic, it is possible to aggregate citations by averages. 
Suppose that a specific technical topic occurred at year $N$ (e.g., Deep Learning at 2012), and we would like to quantify its impact.
One simple method is to compare the number of papers whose topic is neural network before and after year $N$.
However, this method has two problems. 
First, what would have happened without the research is fundamentally a counterfactual phenomenon (counterfactual problem (P1)). 
Second, the difference between before and after a specific technical topic occurred could contain several confounding factors not only the effect of the technical topic occurred but also a temporal trend, etc. This is the well-known problem of pre-post study design (pre-post comparison problem (P2))~\cite{campbell1963experimental,dimitrov2003pretest}.

\begin{figure*}[tb]
  \centering
  \includegraphics[width=\linewidth]{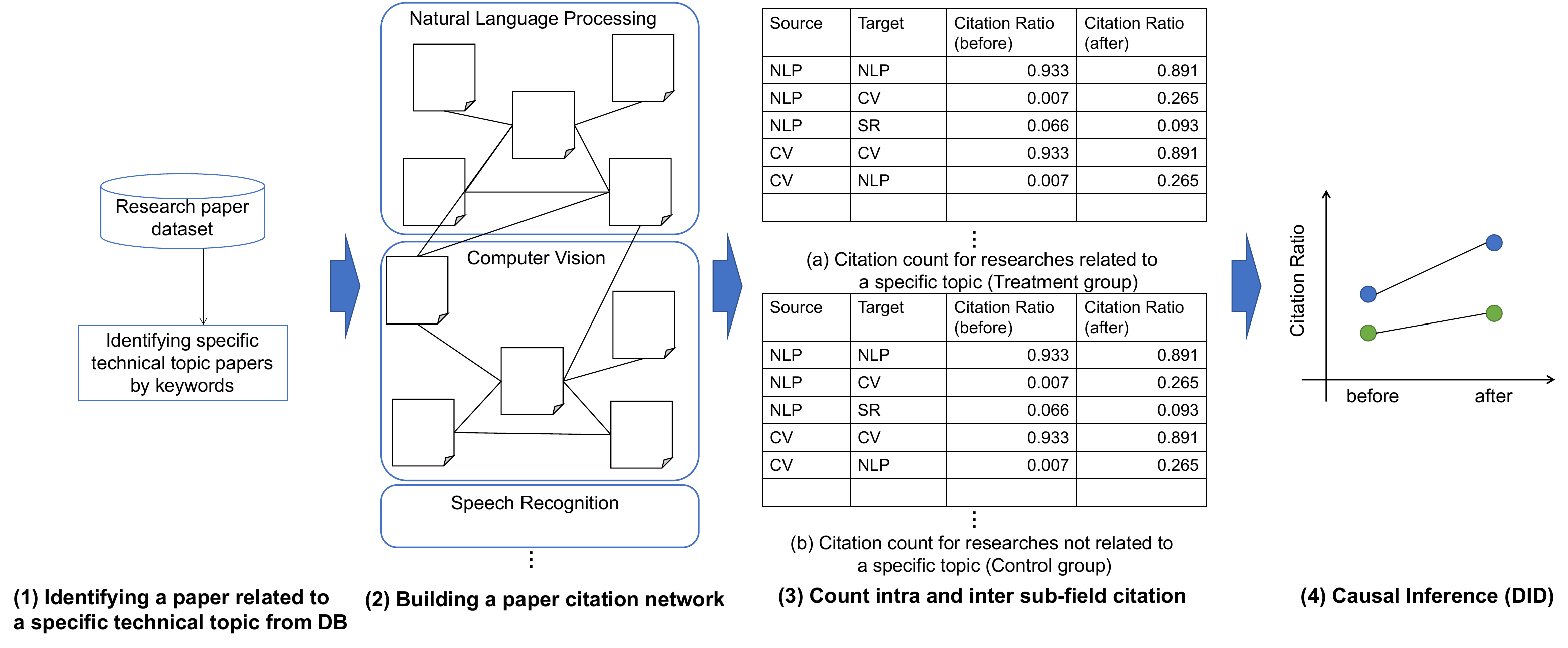}
  \caption{Overview of the proposed approach.}
  \label{fig:overview}
\end{figure*}

% 4. Why hasn't it been solved before? (Or, what's wrong with previous proposed solutions? How does mine differ?)
% 5. What are the key components of my approach and results? Also include any specific limitations.
In this paper, we propose a framework based on causal inference to quantify the research impact of a specific technical topic using a large-scale citation dataset.
The key idea of our approach is leveraging difference-in-difference~\cite{Lechner:2014} to bibliometric data.
Figure \ref{fig:overview} shows the conceptual overview of the proposed approach.
First, we identify papers of a specific technical topic from the research paper dataset using keywords or category tags. The identified papers are referred to as treatment group papers, and the remained papers are referred to as control group papers.
We use Microsoft Academic Graph (MAG) \cite{Sinha:2015,wang2019review,wang2020microsoft} which is one of the largest academic publication dataset as a research paper dataset.
Second, we build a paper citation network using the treatment group papers, the control group papers, and papers cited by both groups of papers. 
Next, the number of citations related to a specific topic (treatment group) and those not related to a specific topic (control group) is counted for each research field. 
Finally, the impact of a research topic for each research field is estimated by applying difference-in-difference. Exploiting causal inference can solve the counterfactual problem (P1) and pre-post comparison problem (P2).
We validate our framework using several technical topics (e.g., deep learning and interpretability for ML models) for AI top conferences including quantifying the research impact of deep learning, and comparing the research impact of deep learning with other topics.
Evaluation results show that (1) deep learning significantly affects computer vision (+42.15\% publication) and natural language processing (+35.31\% publication),
(2) data mining papers cite many papers in various fields, (3) deep learning significantly affect cross-field citation especially for speech recognition to computer vision (+17.4\% p.p.), and natural language processing to computer vision (+7.1\% p.p.), and (4) the quantified research impact indicated that the effect of deep learning was 3.1 times of the effect of interpretability for ML models.

The contributions of this study are the following:
\begin{itemize}
  \item We propose a conceptual framework based on causal inference to quantify the research impact.
  \item We present the case studies (1) to quantify the research impact of deep learning, and (2) to compare the several technical topics by exploiting Microsoft Academic Graph.
  \item Evaluation results show that deep learning significantly affects computer vision and natural language processing.
\end{itemize}

The rest of this paper is organized as follows. 
In the next section, we reviews related work on the analysis of bibliometric data.
We explain the dataset used in this study in Section \ref{sec:data}.
Section \ref{sec:approach} describes the proposed approach.
We evaluate our approach in Section \ref{sec:eval}.
Finally, we conclude this study and discuss future work.

\section{Related Work}\label{sec:related_work}
\subsection{Quantifying research impact}\label{sec:related_work_quantify}
Most basic approach to quantify a research impact is to use citation count, such as total number of citations and average number of citations per publication~\cite{waltman2016review}.
However, it is necessary to calculate the number of citations by considering various factors (e.g., time window and research field) depending on the purpose of comparing studies.
In the existing studies, several types of research impacts have been proposed such as long-term impact~\cite{wang2013quantifying}, impact of individual~\cite{sinatra2016quantifying}
and impact scores for individual research publications~\cite{sutherland2011quantifying}.
In these studies, the situation of what would have happened without the study is not taken into account.

There are several studies which exploit causal inference for quantifying the research impact. 
Farys and Wolbring estimated the effect of the Nobel Prize for publication using matched control group~\cite{farys2017matched}.
Bornmann et al. proposed to leverage the propensity score to consider the difference in the number of publications in each field (i.e., field-normalization)~\cite{bornmann2020should}.
Since DID is more suitable for the estimation considering the fixed effect than propensity score matching~\cite{gebel2014impact}, we proposed to leverage the DID method.

\subsection{Citation Prediction and Trend Detection}\label{sec:related_work_prediction}
Citation prediction and trend detection are widely studied research topics~\cite{yu2012citation,pobiedina2016citation,cao2016data}.
For example, Abrishami et al. proposed a method to predict citation count using deep neural networks~\cite{abrishami2019predicting}.
Dong et al. predicted the effect for the author's {\it h}-index by publishing a paper~\cite{dong2015will}.
Asatani et al. proposed a framework that detects the research trend using network representation learning for a citation network~\cite{asatani2018detecting}.
Recently, Shao et al. investigated the evolution and trend of artificial intelligence using an academic knowledge graph on AI, called AI 2000~\cite{Shao:2020}.
Besides, Frank et al. \cite{frank2019evolution} analyzed the evolution of AI.
These studies focus on prediction and detection method which differs from our study.

\begin{table*}[tb]
\caption{List of AI related conferences}
\label{tab:conf_list}
\centering
\begin{tabular}{lll}
\toprule
Conference Name                                                           & Acronym     & Sub-field \\ \midrule
AAAI Conference on Artificial Intelligence                          & AAAI        & Classical AI        \\
International Conference on Artificial Intelligence and Statistics      & AISTATS     & Classical AI        \\
Conference on Learning Theory                                             & COLT        & Classical AI        \\
European Conference on Artificial Intelligence                          & ECAI        & Classical AI        \\
International Joint Conference on Artificial Intelligence               & IJCAI       & Classical AI        \\
Uncertainty in Artificial Intelligence                                    & UAI         & Classical AI        \\
Computer Vision and Pattern Recognition                                   & CVPR        & CV        \\
European Conference on Computer Vision                                    & ECCV        & CV        \\
International Conference on Computer Vision                             & ICCV        & CV        \\
International Conference on Data Mining                                   & ICDM        & DM        \\
Knowledge Discovery and Data Mining                                       & KDD         & DM        \\
SIAM International Conference on Data Mining                            & SDM         & DM        \\
Web Search and Data Mining                                                & WSDM        & DM        \\
International Conference on Learning Representations                    & ICLR        & ML        \\
International Conference on Machine Learning                            & ICML        & ML        \\
Annual Conference on Neural Information Processing Systems                                     & NeurIPS     & ML        \\
Meeting of the Association for Computational Linguistics                & ACL         & NLP       \\
International Conference on Computational Linguistics                   & COLING      & NLP       \\
Empirical Methods in Natural Language Processing                        & EMNLP       & NLP       \\
North American Chapter of the Association for Computational Linguistics & NAACL       & NLP       \\
International Conference on Acoustics, Speech, and Signal Processing    & ICASSP      & SR        \\
Conference of the International Speech Communication Association        & INTERSPEECH & SR        \\ \bottomrule
\end{tabular}
\end{table*}

\section{Data}\label{sec:data}
We use Microsoft Academic Graph (MAG) \cite{Sinha:2015,wang2019review,wang2020microsoft} which is one of the largest academic publication dataset provided by Microsoft\footnote{https://www.microsoft.com/en-us/research/project/microsoft-academic-graph/}.
MAG contains paper information such as paper title, author, published year, venue, field of study (technical topic).
Comprehensive data schema is provided at the web page of MAG\footnote{https://docs.microsoft.com/en-us/academic-services/graph/reference-data-schema}.
As of February 1, 2021, MAG consists of 251,493,799 papers and 53,529 venues.
To focus on AI field, we filter the dataset by AI related venues.
To select the venues, we follow the work of Shao et al.~\cite{Shao:2020} which defines 20 sub-fields of AI such as Classical AI, Machine Learning (ML), Data Mining (DM), Computer Vision (CV), Natural Language Processing (NLP), Speech Recognition (SR) etc. They made these sub-field by interviewing experts in each sub-field. 
We slightly change the list of venues for each sub-field.
For example, COLT (Conference on Learning Theory), which is originally assigned to NLP, is changed to an AI conference rather than an NLP conference.
The complete venue lists of each sub-field are shown in Table \ref{tab:conf_list}.

\begin{figure}[tb]
  \centering
  \includegraphics[width=\linewidth]{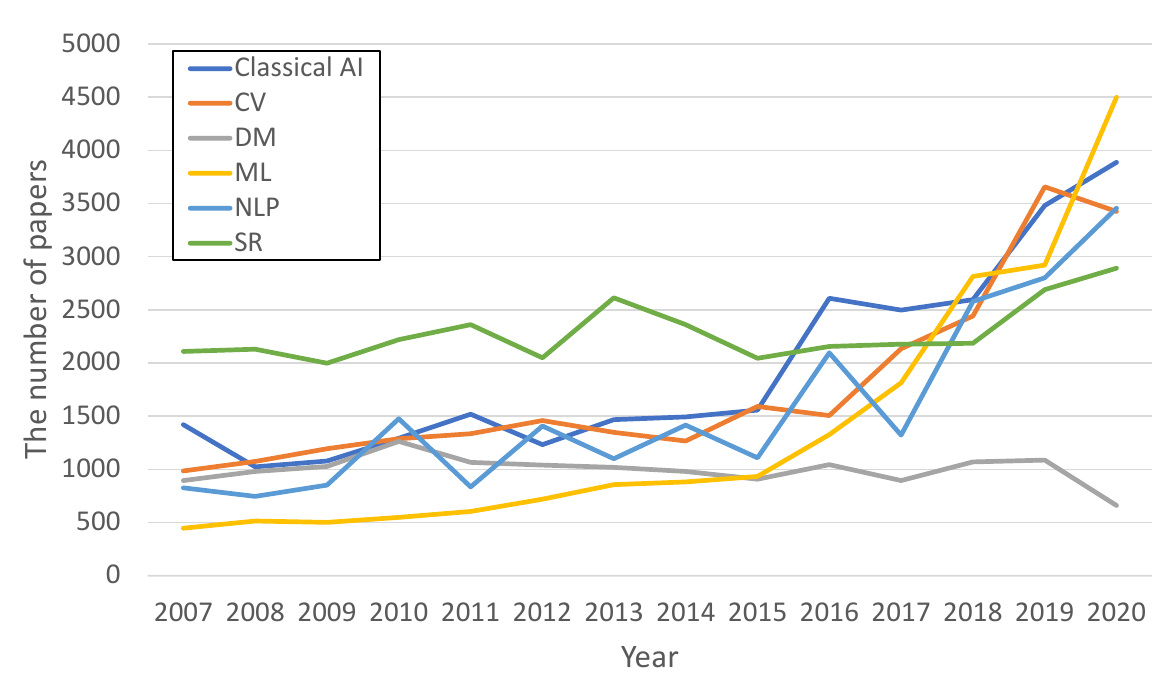}
  \caption{The number of papers for each research field from 2007 to 2020.}
  \label{fig:num_papers}
\end{figure}

Figure \ref{fig:num_papers} shows the number of published papers for each research field from 2007 to 2020. Because the transition of the number of papers has an uptrend, we have to consider the trend to compare the data of different time points.
Finally, the number of selected venues is 39, and the number of papers is 318,061.

\section{Proposed Framework}\label{sec:approach}
Our goal is to quantify the impact of the emergence of a specific technical topic, taking causality into account.
The ideal method is to perform a so-called Randomized Controlled Trial (RCT), in which research topics in each paper are randomly assigned and compared the citation count.
However, this is impractical, as researchers cannot be forced to write a paper regarding a specific topic.
Thus, we adopt quasi-experimental design \cite{cook2002experimental} 
which assesses a causality based purely on the observational data.
Specifically, we conduct difference-in-differences (DID) \cite{Lechner:2014}, commonly used in economics \cite{donald2007inference} and healthcare \cite{dimick2014methods} literature.
Similar approach is used to quantify the effect of audio streaming consumption (i.e., music vs podcast) \cite{Li:2020}, the effect of online social actions to offline behavior\cite{Althoff:2017} and the change of human needs during COVID-19 pandemic using web search logs\cite{Suh:2021}.
We treat the emergence of a novel specific technical topic as the intervention.
Then, papers which are related to a specific technical topic are considered as treatment group, 
and papers which are not related to a specific technical topic are considered as control group.

To conduct DID, the proposed framework consists of four steps.
\begin{enumerate}
  \item Identifying treatment group papers (paper related to a specific technical topic) from academic paper database.
  \item Building a paper citation network using the treatment group papers and papers cited by the treatment group papers
  \item Counting intra and inter sub-field citation using the paper citation network.
  \item Conducting DID to estimate the effect of a specific technical topic emerged.
\end{enumerate}

\subsection{Identifying treatment group papers}\label{sec:identify_papers}
A naive method to identify a technical topic of a paper is keyword matching to paper title or abstract, whereas MAG dataset has more rich information regarding technical topic of a paper.
MAG contains {\it field of study} field which represents the technical topic of each paper such as deep neural networks, gradient boosting etc.
Thus, we exploit field of study to identify the technical topic of a paper.
Field of study for each paper is labeled automatically \cite{Sinha:2015}, and the topical hierarchy is generated by applying hierarchical topic modeling \cite{shen2018web}.

To create the treatment group papers which are related to a specific topic, we extract the papers by keyword matching using list of keywords for field of study field.
For example, if we would like to extract papers related to deep learning, the examples of keyword are {\it neural network}, {\it deep learning}, and {\it long short term memory}.
In out experiment, we further filter the papers by AI related conferences defined in \ref{sec:data} to focus on AI related topics, and added sub-fields of AI (Classical AI, ML, DM, CV, NLP, SR).
The remained papers, which are not matched to the keyword list in field of study field, are defined as control group papers.
Finally, both the treatment group papers and the control group papers contain paper id, published year, venue, field of study, sub-fields of AI.

\subsection{Building a paper citation network}\label{sec:build_nw}
In this step, we first extract papers which are cited by the treatment group papers from MAG, called as cited papers for treatment.
Then, a paper citation network is built using the treatment group papers and cited papers for treatment.
Similar to treatment group, cited papers for control group papers are also extracted from MAG, and a paper citation network for control group papers is build using the control group papers and cited paper for control group.
At the end of this step, the processed dataset contains original papers and cited papers information (paper id, venue, published year, sub-fields of AI for both original papers and cited papers, and reference relationship).

\subsection{Calculating paper citation}\label{sec:count_citation}
Most simple method to calculate the citation measure is simply count the number of paper 
\begin{equation}
C(T,o_{group}) = paper\_num_{T,o}\label{eq:citation_count}
\end{equation}
where $T=[t_s, t_e]$ is target period ($t_s$ is start time and $t_e$ is end time),
$o_{group}$ is the research field of the original paper, $group \in {treatment, control}$,
and $paper\_num_{T,o}$ is the number of papers of field $o$ within time period $T$.

One of our objective is to quantify the effect of the emergence of a technical topic to other research field.
Thus, we can also consider the inter-field citation.
Because the number of papers varies depending on the research field, 
field-normalization~\cite{waltman2016review} should be applied to remove a selection bias.
Hence, we calculate the citation ratio for comparing the inter-field citation effect as follows.
\begin{equation}
C(T,o_{group},d) = \frac{paper\_num_{T,o,d}}{paper\_num_{T,o}}\label{eq:citation_ratio}
\end{equation}
where $d$ is the research field of cited paper,
and $paper\_num_{T,o,d}$ is the number of papers cited from field $o$ to field $d$ within time period $T$.
Field $o$ and $d$ is in $\{Classical\ AI, ML, DM, CV, NLP, SR\}$.
For example, if $T=[2013,2017]$, $o_{group}=CV_{treatment}$, $d=NLP$, and $treatment$ is a group of papers whose technical topic is deep learning,
then $CR(T,o_{group},d)$ represents the percentage of papers that cite NLP papers from CV papers of deep learning from 2013 to 2017 for treatment group (a specific topic).

\subsection{Difference-in-differences}\label{sec:DID}
Before conducting DID, we need to confirm that the parallel trends assumption is met.
There are several method to verify the parallel trends assumption.
We adopt the parallel trend test using slope of regression model before the intervention \cite{muralidharan2017cycling,akosa2013effects}.
Specifically, we test the difference of the slopes for regression models of both treatment group and control group \cite{armitage2008statistical}.
As an alternative method, we can adopt propensity score matching \cite{Li:2020} to validate the indistinguishable of two groups.

Then, the average treatment effect (ATE) is estimated by DID by comparing the average difference of the before and after of each group.
There are two methods to calculate ATE, i.e., absolute effect and relative effect.
The absolute effect can be calculated as follows.
\begin{equation}
\begin{split}
    ATE(T_1,T_2,o,d) = (C(T_2,o_{treatment},d) - C(T_1,o_{treatment},d)) \\
    - (C(T_2,o_{control},d) - C(T_1,o_{control},d)) \label{eq:DID_absolute}
\end{split}
\end{equation}

On the other hand, the relative effect can be calculated similar to the work of \cite{Suh:2021} as follows.
\begin{equation}
\begin{split}
    ATE(T_1,T_2,o,d) = \log_2\left(
\frac{C(T_2,o_{treatment},d)}{C(T_1,o_{treatment},d)}\right) \\
     - \log_2\left(\frac{C(T_2,o_{control},d)}{C(T_1,o_{control},d)}\right) \label{eq:DID_relative}
\end{split}
\end{equation}

\section{Analyses and Results}\label{sec:eval}
In this section, we conduct a qualitative evaluation through case studies by considering research questions.

\subsection{Case 1: Deep Learning}\label{sec:case1}
\noindent{\bf Background: }
One of the important role of deep learning is representation learning that allows a machine to automatically generate the representations~\cite{lecun2015deep}, not the hand-crafted representations, needed for classification or regression.
While hand-crafted feature was designed by using domain knowledge in each research field before the rise of deep learning, the realization of representation learning using deep learning is thought to have accelerated research across fields. 
Therefore, we aim to quantify the impact of deep learning as a technical field in the research of computer science domain.
Specifically, we solve the following research questions.
\begin{itemize}
    \item {\bf RQ1: How much did deep learning affect each field of computer science?}
    \item {\bf RQ2: To what extent does deep learning impact the integration of computer science disciplines?}
\end{itemize}

To answer these research questions, we used the following setting.
To identify the research related to deep learning (which indicates that the paper is regarded as the treatment group), 
the paper which included the word of the {\it neural network} in the field of study was extracted.
We set the time of intervention was 2012 because the drastic improvement of image recognition accuracy in ILSVRC 2012 was one of the opportunities of the popularization of deep learning~\cite{krizhevsky2012imagenet}.
Thus, we set $T_1=[2007, 2011]$ and $T_2=[2013, 2017]$ (i.e., before and after 5 years).
We count the number of paper citation by Eq.(\ref{eq:citation_ratio}).
We estimated the ATE by Eq.(\ref{eq:DID_relative}) for RQ1, and Eq.(\ref{eq:DID_absolute}) for RQ2.

\subsubsection{RQ1: How much did deep learning affect each field of computer science?} 
\begin{table}[tb]
\caption{Relative change in the number of publication in each research field for deep learning.}
\label{tab:result_DNN}
\centering
\begin{tabular}{lr}
\toprule
Research Field & \multicolumn{1}{l}{Relative Change (\%)} \\
\midrule
Classical AI             & 5.05                                     \\
Computer Vision             & 42.15                                    \\
Data Mining             & 3.84                                     \\
Machine Learning             & 5.76                                     \\
Natural Language Processing            & 35.31                                    \\
Speech Recognition             & 6.02                                    \\
\bottomrule
\end{tabular}
\end{table}

Table \ref{tab:result_DNN} shows the result of relative change in the number of publication in each research field for deep learning. Note that when we test the parallel trend for pre-intervention period from 2002 to 2011, most research fields except for Classical AI were not met the parallel trend. 
Although we need to adjust covariates from this results, we can see that the number of publication regarding deep learning has greatly increased after 2012, especially for computer vision and natural language processing.

\subsubsection{RQ2: To what extent did deep learning impact the integration of computer science disciplines?} 
\begin{table*}[tb]
\caption{Change of percentage points of inter-field citation in each research field for deep learning.}
\label{tab:result_DNN_inter}
\centering
\scalebox{0.8}{
\begin{tabular}{llrlllr}
\cline{1-3} \cline{5-7}
Original Field  & Cited Field                 & \multicolumn{1}{l}{Change of p.p.} &  & Original Field              & Cited Field                 & \multicolumn{1}{l}{Change of p.p.} \\ \cline{1-3} \cline{5-7} 
Classical AI    & Classical AI                & -1.6                                     &  & Machine Learning            & Classical AI                & -10.0                                    \\
Classical AI    & Computer Vision             & 3.6*                                     &  & Machine Learning            & Computer Vision             & 6.7*                                     \\
Classical AI    & Data Mining                 & 2.8                                      &  & Machine Learning            & Data Mining                 & 1.7                                      \\
Classical AI    & Machine Learning            & -2.0                                     &  & Machine Learning            & Machine Learning            & -1.1                                     \\
Classical AI    & Natural Language Processing & -1.8                                     &  & Machine Learning            & Natural Language Processing & 9.1                                      \\
Classical AI    & Speech Recognition          & 5.6                                      &  & Machine Learning            & Speech Recognition          & -6.9                                     \\
Computer Vision & Classical AI                & -13.3                                    &  & Natural Language Processing & Classical AI                & 0.9                                      \\
Computer Vision & Computer Vision             & 3.3                                      &  & Natural Language Processing & Computer Vision             & 7.1*                                     \\
Computer Vision & Data Mining                 & 2.4                                      &  & Natural Language Processing & Data Mining                 & 3.0                                      \\
Computer Vision & Machine Learning            & -4.8                                     &  & Natural Language Processing & Machine Learning            & -19.7                                    \\
Computer Vision & Natural Language Processing & -0.1                                     &  & Natural Language Processing & Natural Language Processing & -1.2                                     \\
Computer Vision & Speech Recognition          & 0.5                                      &  & Natural Language Processing & Speech Recognition          & -29.8                                    \\
Data Mining     & Classical AI                & 33.0                                     &  & Speech Recognition          & Classical AI                & -2.2*                                    \\
Data Mining     & Computer Vision             & 22.4                                     &  & Speech Recognition          & Computer Vision             & 17.4                                     \\
Data Mining     & Data Mining                 & 11.7                                     &  & Speech Recognition          & Data Mining                 & 0.4*                                     \\
Data Mining     & Machine Learning            & 20.4                                     &  & Speech Recognition          & Machine Learning            & 2.0*                                     \\
Data Mining     & Natural Language Processing & 23.3                                     &  & Speech Recognition          & Natural Language Processing & 1.1*                                     \\
Data Mining     & Speech Recognition          & 9.9                                      &  & Speech Recognition          & Speech Recognition          & -10.6*                                   \\ \cline{1-3} \cline{5-7} 
\end{tabular}
}
\end{table*}

Table \ref{tab:result_DNN_inter} shows the inter-field effect of deep learning.
{\it Original Field} indicates the paper cite other field, {\it Cited Field} indicates the paper cited from other research fields, {\it Change of p.p.} represents the change of percentage points of citation (CoP for short),
and $*$ means that the parallel trend was not met.
From this table, the following trends are observed by focusing on CoP increases of more than 10 points.
\begin{enumerate}
    \item Data mining papers cite many papers in various fields.
    \item Speech recognition papers cite computer vision papers.
\end{enumerate}
For (1), it seems natural that data mining utilizes technologies in various fields.
For (2), deep neural networks were used in speech recognition in 2012~\cite{hinton2012deep}, and it seems to have been affected by deep learning from an early stage. 
Besides, because the paper of ICASSP 2013~\cite{deng2013new}\footnote{The number of citations is 1060 as of Oct, 17th, 2021 by Google scholar} by Deng, Hinton and Kingsbury noted that 
``Convolutional neural networks have been widely used in computer vision where they have been very successful \cite{krizhevsky2012imagenet}'',
it may have triggered the citation of the image recognition field.

\subsection{Case 2: Comparison of deep learning with other technical topics}\label{sec:case2}
\noindent{\bf Background: }
Deep learning is not the only technology at the top of the AI conference. 
For example, gradient boosting decision tree (GBDT) such as LightGBM~\cite{ke2017lightgbm} and XGBoost~\cite{chen2016xgboost}, and the interpretability of ML models such as LIME~\cite{ribeiro2016should} and SHAP~\cite{lundberg2017unified} are also attracting attention.
Thus, we aim to quantitatively compare the impact of deep learning with other technical topics in computer science.
Specifically, we solve the following research questions.
\begin{itemize}
    \item {\bf RQ3: To what extent did deep learning affect other technical topics?}
\end{itemize}

To answer this research question, we used the following setting.
For deep learning, we used the same setting in the previous subsection.
To identify the research related to GBDT, the paper which included the word of the {\it gradient boost} in the field of study was extracted.
Similarly, to identify the research related to interpretability of ML models,
the paper which included the word of the {\it interpretability} in the field of study was extracted.

We set the time of intervention for GBDT was 2016 because the paper of XGBoost was published at KDD 2016. 
Similarly, the time of intervention for interpretability of ML models was set at 2016 because the paper of LIME~\cite{ribeiro2016should} was published at KDD 2016.
Thus, we set $T_1=[2010, 2015]$ and $T_2=[2016, 2020]$ (i.e., before and after 5 years) for both GBDT and interpretability of ML models.
We count the number of paper citation by Eq.(\ref{eq:citation_ratio}).
and the ATE is estimated using Eq.(\ref{eq:DID_absolute}).

\subsubsection{RQ3: To what extent did deep learning affect other technical topics?} % 

\begin{table}[tb]
\caption{Comparison of change of percentage of inter-field citation in each research field.}
\label{tab:result_comparison}
\centering
\scalebox{1.0}{
\begin{tabular}{lrrr}
\toprule
\multirow{2}{*}{Research Field} & \multicolumn{3}{c}{Relative Change (\%)}                                                            \\ \cline{2-4} 
                                & \multicolumn{1}{l}{DL} & \multicolumn{1}{l}{GBDT} & \multicolumn{1}{l}{INTP} \\ \midrule
Classical AI                    & 5.05                              & -0.19                    & 1.54                                 \\
Computer Vision                 & 42.15                             & -0.56                    & 17.44                                \\
Data Mining                     & 3.84                              & -0.42                    & 1.91                                 \\
Machine Learning                & 5.76                              & 0.14                     & 1.46                                 \\
Natural Language Processing     & 35.31                             & 3.35                     & 4.09                                 \\
Speech Recognition              & 6.02                              & 4.05                     & 5.14                                 \\ \hline
Average                         & 16.35                             & 1.06                     & 5.26                                 \\ \bottomrule
\end{tabular}
}
\end{table}

Table \ref{tab:result_comparison} shows the result of different technical topics.
DL indicates deep learning, GBDT indicates gradient boosting decision tree, and INTP indicates interpretability of ML models.
Because the order of the research impact of each technology was deep learning, interpretability of ML models and GBDT by comparing the average of the relative change,
the research impact of deep learning is the most influential.
Here, we discuss these results in more depth. 
Notably, NLP and SR have a relatively high values of the relative change compared to other research field for GBDT. 
In NLP and SR, a tree-based classifier such as GBDT may be likely to be used because the structure of sentence modification is handled in a tree structure.
Moreover, CV has a significantly high value of the relative change compared to other research field. 
In CV field, the relative change may be high because studies on the interpretability peculiar to images such as Grad-CAM~\cite{selvaraju2017grad} are being conducted.

\subsection{Limitations and Implications}\label{sec:discussion}
\noindent{\bf Limitation:}
Our work has the following limitations.\newline
(1) Method of causal inference: Although we used the DID method, if concurrent technical trends have significantly influenced the number of citation, the reliability of the estimated impact may be low. 
Besides, because the parallel trend was not met in several cases, if we adopt propensity score matching, the estimated results can be more reliable.
Futhermore, there may be potentially spill-over effects between the treatment and control groups (e.g., the citations between treatment and control papers). In this case, the vanilla DID is not suitable for this analysis.

\noindent{(2) Identifying a technical topic:} in our implementation, the identification of a technical topic relies on the {\it field of study} field automatically tagged by the work of \cite{Sinha:2015}.
Thus, incorrect tagging leads to incorrect estimation.
We can alternatively identify the topic of a paper by using keyword matching for the title or abstract. Then, the estimated impact may be affected.

\noindent{\bf Implications:}
Our results lead to implications for the prediction of research impact.
If we combine the proposed research impact measure and the paper citation prediction,
we can obtain more precise research impact on each research field.
Besides, if keywords representing research topics are prepared in an exhaustive manner, it is possible to quantitatively compare the impact of each research topic.
The government can use the estimation results to make decisions on the selection of research areas for investment.

\section{Conclusion}
We have proposed a conceptual framework based on causal inference to quantify the research impact of a specific technical topic using large-scale citation dataset.
The proposed framework exploited the difference-in-difference to quantify the research impact.
We validated our framework using several technical topic (e.g., deep learning and interpretability for ML models) for AI top conferences through two case studies. 
Our findings was (1) computer vision and natural language processing were the most influenced research field by deep learning (+42.15\% publication for computer vision, and +35.31\% publication for natural language processing), 
(2) data mining research exploited the research of various fields, 
(3) deep learning significantly promoted cross field citation especially for speech recognition to computer vision (+17.4\% p.p.), and natural language processing to computer vision (+7.1\% p.p.),
and (4) the impact of deep learning was 3.1 times of the impact of interpretability for ML models.
We believe that this study opens up a novel bibliometrics approach for quantifying the research impact, and hope that the approach will be helpful in real world.

\begin{acks}
We would like to thank Microsoft for providing Microsoft Academic Graph dataset.
\end{acks}

\bibliographystyle{ACM-Reference-Format}
\bibliography{reference_DL}

% \appendix

% \section{List of AI related conferences}
% Table \ref{tab:conf_list} shows all list of conferences used in our evaluation.

\end{document}